\documentclass[11pt]{bmc_article_s50}

\usepackage{booktabs}

\usepackage{multirow}
\usepackage{color}
\usepackage{colortbl,rotating}
\usepackage{hhline}

\usepackage{url}
\usepackage[utf8]{inputenc}

\usepackage{url}
\usepackage{setspace}
\usepackage[T1]{fontenc}
\usepackage{ragged2e}
\justifying
\singlespace

\begin{document}

\title{Considering human aspects on strategies for designing and managing distributed human computation}

\maketitle

\author[1^{*}]{Lesandro Ponciano}\cor{}
\email{lesandrop@lsd.ufcg.edu.br}

\author[1]{Francisco Brasileiro}
\email{fubica@computacao.ufcg.edu.br}

\author[1]{Nazareno Andrade}
\email{nazareno@computacao.ufcg.edu.br}

\author[1]{L\'{i}via Sampaio}
\email{livia@computacao.ufcg.edu.br}

\address[1]{Federal University of Campina Grande, Department of Computing and Systems, Av.\\\hspace*{59pt}  Apr\'{i}gio Veloso, 882 -- Bloco CO, 58.429-900, Campina Grande -- PB, Brazil}

\begin{abstract}\justifying
A human computation system can be viewed as a distributed system in which the processors are humans, called workers. Such systems harness the cognitive power of a group of workers connected to the Internet to execute relatively simple tasks, whose solutions, once grouped, solve a problem that systems equipped with only machines could not solve satisfactorily. Examples of such systems are Amazon Mechanical Turk and the Zooniverse platform. A human computation application comprises a group of tasks, each of them can be performed by one worker. Tasks might have dependencies among each other. In this study, we propose a theoretical framework to analyze such type of application from a distributed systems point of view. Our framework is established on three dimensions that represent different perspectives in which human computation applications can be approached: quality-of-service requirements, design and management strategies, and human aspects. By using this framework, we review human computation in the perspective of programmers seeking to improve the design of human computation applications and managers seeking to increase the effectiveness of human computation infrastructures in running such applications. In doing so, besides integrating and organizing what has been done in this direction, we also put into perspective the fact that the human aspects of the workers in such systems introduce new challenges in terms of, for example, task assignment, dependency management, and fault prevention and tolerance. We discuss how they are related to distributed systems and other areas of knowledge.


\end{abstract}

Journal of Internet Services and Applications 2014, 5:10, Springer London, ISSN: 1867-4828 \\ http://dx.doi.org/10.1186/s13174-014-0010-4 

\section{Introduction}

Many studies have focused on increasing the performance of machine-based computational systems over the last decades. As a result, much progress has been made allowing increasingly complex problems to be efficiently solved. However, despite these advances, there are still tasks that cannot be accurately and efficiently performed even when the most sophisticated algorithms and computing architectures are used~\cite{Quinn2011,Savage2012}. Examples of such tasks are those related to natural language processing, image understanding and creativity~\cite{Yuen:2009,King2011}. A common factor in these kinds of tasks is their suitability to human abilities; human beings can solve them with high efficiency and accuracy~\cite{Quinn2011,Savage2012}. In the last years, there has emerged a new computing approach that takes advantage of human abilities to execute these kinds of tasks. Such approach has been named \textit{Human Computation}~\cite{Quinn2009,Quinn2011}.

Applications designed to execute on human computation systems may encompass one or multiple tasks. They are called \textit{distributed human computation applications} when they are composed of multiple tasks, and each individual task can be performed by a different human being, called \textit{worker}. In the last years, distributed computing systems have been developed to support the execution of this type of application. They gather a crowd of workers connected to the Internet and manage them to execute application tasks. The precursor of such systems is reCAPTCHA~\cite{von:Science:2008}. Currently, there is a broad diversity of distributed human computation applications and distributed systems devoted to execute them, such as: games with a purpose~\cite{vonAhn:2008}, contests sites~\cite{Archak:2010}, online labor markets~\cite{Ipeirotis:2010}, and volunteer thinking systems~\cite{Ponciano:CiSE:2014}. In this paper, we focus on online labor markets and volunteer thinking systems.

Online labor markets gather a crowd of workers that have a financial motivation~\cite{Ipeirotis:2010}. The precursor of this type of system is the Amazon Mechanical Turk platform (mturk.com). Such plaform reports to have more than $400,000$ registered workers~\cite{Ross:2010}, and receives between $50,000$ and $400,000$ new tasks to be executed per day (mturk-tracker.com) at the time of writing. Volunteer thinking systems, in turn, gather a crowd of workers willing to execute tasks without any financial compensation~\cite{Ponciano:CiSE:2014}. One of the precursors of this type of system is the Zooniverse citizen-science platform (zooniverse.org). Currently, Zooniverse hosts $21$ scientific projects and has over one million registered workers. Only Galaxy Zoo, the largest project at Zooniverse, had $50$ million tasks performed by $150,000$ workers in a year of operation~\cite{ball:2013}. Thus, both labor markets and volunteer thinking are large-scale distributed human computation systems.

Because the computing units in human computation systems are human beings, both the design and management of applications tap into concepts and theories from multiple disciplines. Quinn and Bederson conducted one of the first efforts to delimit such concepts and theories~\cite{Quinn2011,Quinn2009}. They present a taxonomy for human computation, highlighting differences and similarities to related concepts, such as collective intelligence and crowdsourcing. Yuen et al., in turn, focus on distinguishing different types of human computation systems and platforms~\cite{Yuen:2009,King2011}. More recently, Kittur et al. built a theoretical framework to analyze future perspectives in developing online labor markets that are attractive and fair to workers~\cite{Kittur:cscw:2013}. Differently from previous efforts, in this study, we analyze human computation under the perspective of \textit{programmers seeking to improve the design of distributed human computation applications} and \textit{managers seeking to increase the effectiveness of distributed human computation systems}.

To conduct this study, we propose a theoretical framework that integrates theories about human aspects, design and management (D\&M) strategies, and quality of service (QoS) requirements. Human aspects include characteristics that impact workers' ability to perform tasks (e.g., cognitive system and emotion), their interests (e.g., motivation and preferences), and their differences and relations (e.g., individual differences and social behavior). QoS requirements, in turn, are metrics directly related to how application owners measure applications and systems effectiveness. These metrics are typically defined in terms of time, cost, fidelity, and security. Finally, D\&M strategies consist of strategies related to how the application is designed and managed. They involve activities such as application composition, task assignment, dependency management, and fault prevention and tolerance.

This framework allows us to perform a literature review that expands previous literature reviews to build a vision of human computation focused on distributed systems issues. We emphasize our analysis on three perspectives: $1)$ findings on relevant human aspects which impact D\&M decisions; $2)$ major D\&M strategies that have been proposed to deal with human aspects; and $3)$ open challenges and how they relate to other disciplines. Besides providing a distributed systems viewpoint of this new kind of computational system, our analysis also puts into perspective the fact that human computation introduces new challenges in terms of effective D\&M strategies. Although these challenges are essentially distributed systems challenges, some of them do not exist in machine-based distributed systems, as they are related to human aspects. These challenges call for combining distributed systems design with theories and mechanisms used in other areas of knowledge where there is extensive theory on treating human aspects, such as Cognitive Science, Behavioral Sciences, and Management Sciences.

In the following, we briefly describe the human computation ecosystem addressed in this paper. Then, we present our theoretical framework.  After that, we analyze the literature in the light of our framework. This is followed by the discussion of challenges and perspectives for future research. Finally, we present our conclusions.

\section{Distributed human computation ecosystem}
\label{sec:bgrw}

The core agents in a distributed human computation ecosystem are: requesters, workers, and platform. \textit{Requesters} act in the system by submitting human computation \textit{applications}. An application is a set of \textit{tasks} with or without dependencies among them. Typically, a human computation task consists of some input data (e.g., image, text) and a set of instructions. There are several types of instructions, such as: transcription of an item content (e.g., reCAPTCHA tasks~\cite{von:Science:2008}), classification of an item (e.g., Galaxy Zoo tasks~\cite{Lintott:2008}), generation of creative content about an item~\cite{Araujo:2013}, ranking and matching items~\cite{Marcus:2011}, etc.

\textit{Workers} are the human beings who act as human computers in the system executing one or more tasks. They generate the task output by performing the instructions upon the received items. After executing a task, the worker provides its \textit{output}. The application output is an aggregation of the outputs of all their tasks. In paid systems, when a task is performed, the requester may accept the solution if the task was satisfactorily performed; otherwise, she/he can reject it. Workers are paid only when their solutions are accepted. The receiving of tasks to be executed, the provision of their outputs, and the receiving of the payment for the performed tasks occur via a human computation platform.

The \textit{Platform} is a distributed system that acts as a middleware receiving requester applications and managing the execution of their tasks by the workers. Platforms manage several aspects of tasks execution, such as: providing an interface and language for tasks' specification, performing task replication, maintaining a job board with a set of tasks waiting to be performed, controlling the state of each task from the submission up to its completion. Examples of platforms with such characteristics are the online labor markets Amazon Mechanical Turk (mturk.com) and CrowdFlower (crowdflower.com), and the volunteer thinking systems Zooniverse (zooniverse.org) and CrowdCrafting (crowdcrafting.org). Online labor markets also implement functionalities that allow requesters to communicate with workers that performed their tasks, provide feedback about their outputs, and pay for the tasks performed by them.

Some studies have analyzed human computation ecosystem. In general, they focus mainly on proposing a taxonomy for the area~\cite{Quinn2009,Quinn2011,Yuen:2009,Vukovic:2009} and discussing platform issues~\cite{Dustdar2012,Crouser:2012}. Quinn et al.~\cite{Quinn2009,Quinn2011} propose a taxonomy for human computation that delimits its similarities and differences when compared to others fields based on human work, such as: crowdsourcing, social computing, and collective intelligence. Vukovic and Yuen et al. focus on classifying human computation platforms based on their function (e.g., design and innovation), mode (e.g., competition and marketplace platforms), or algorithms~\cite{Vukovic:2009,Yuen:2009}. Dustdar and Truong~\cite{Dustdar2012} and Crouser and Chang ~\cite{Crouser:2012} focus on hybrid platforms based on the collaboration between machine and human computing units. Dustdar and Truong~\cite{Dustdar2012} focused on strategies to provide machine computation and human computation as a service, using a single interface. Crouser and Chang ~\cite{Crouser:2012} propose a framework of affordances, i.e., properties that are inherent to human and properties that are inherent to machine, so that they complement each other.

Differently from these previous efforts, in the present work we focus on analyzing strategies for designing and managing distributed applications onto human computation platforms. Our main focus is not to survey existing human computation platforms, but to analyze D\&M strategies that have been proposed to be  used in these kind of platforms. Our analysis is based on a theoretical framework built upon theories and concepts from multiple disciplines dealing with (i) human aspects, such as: Motivation Theory~\cite{Maslow:1943}, Self-determination Theory~\cite{Deci:book:1985}, Sense of Community Theory~\cite{mcmillan:JCP:1996}, Human Error Theory~\cite{James:HumanError:1990}, Coordination Theory~\cite{Malone1994} and Human-in-the-Loop literature~\cite{Cranor:UPSEC:2008}, and (ii) applications design, applications management and QoS aspects, such as: the great principles of computing~\cite{Denning2003}; application design methodologies~\cite{Georgakopoulos:1995,Kiepuszewski:2000}; taxonomies for application management in grid computing~\cite{Yu:2005}, web services~\cite{Cardoso:2002}, and organizations~\cite{Kumar:2002}.

\section{Theoretical framework}
\label{sec:arc}

\begin{figure}[!h]
\centering
\includegraphics[width=0.6\textwidth]{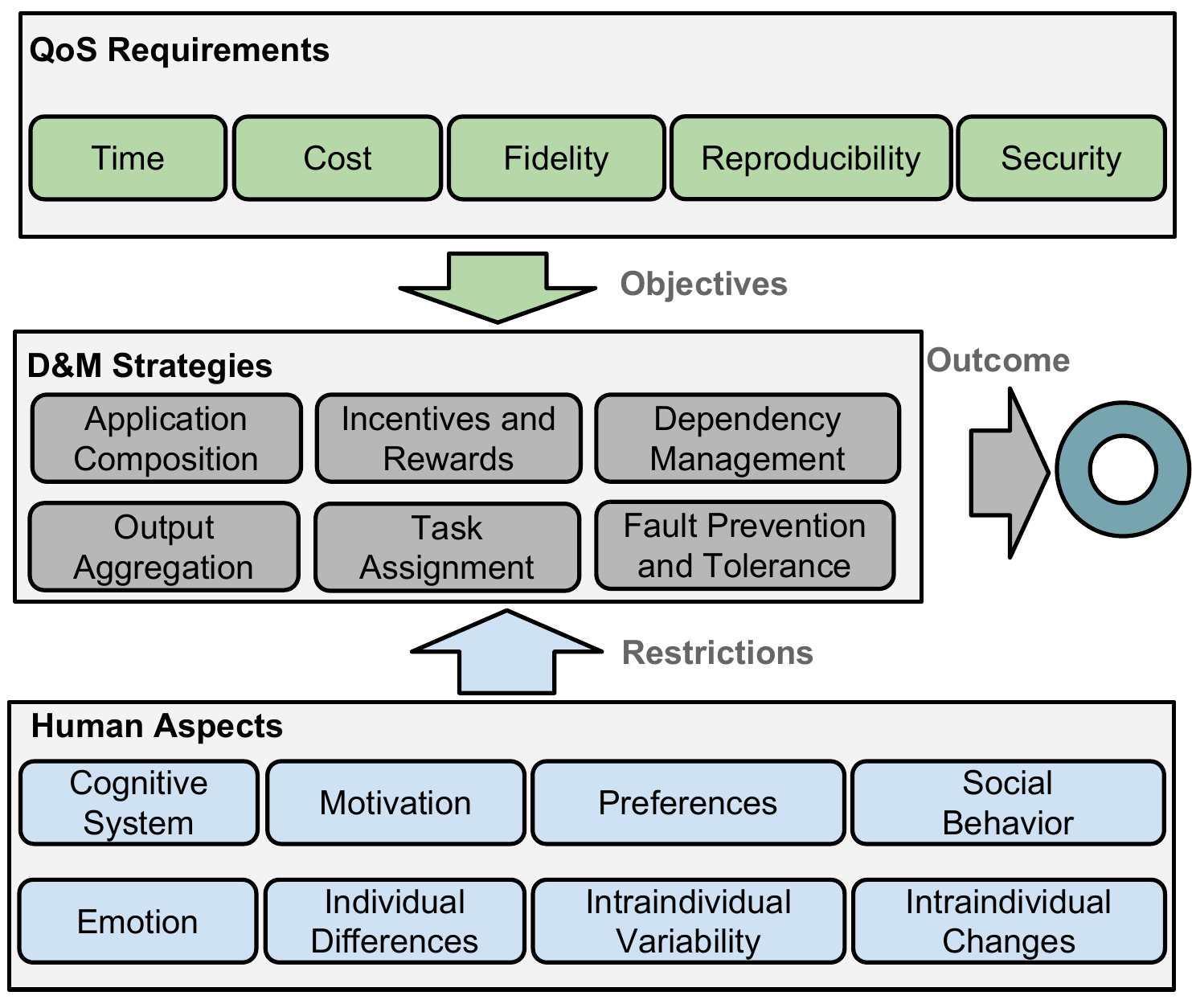}
\caption{Framework.
An overview of the dimensions and factors of the proposed theoretical framework.}
\label{fig1}
\end{figure}

Theoretical frameworks have several distinct roles~\cite{Grudin:2012}. Most important for us, they allow researchers to look for patterns in observations and to inform designers of relevant aspects for a situation. Our framework is designed to assist the analysis of the diverse aspects related to human computation applications. It is organized in three dimensions which represent different perspectives in which it is possible to approach human computation: \textit{QoS requirements}, $D\&M$ \textit{strategies}, and \textit{human aspects}. Each dimension is closely connected to an agent in the human computation ecosystem: QoS requirements are requesters' effectiveness measures; D\&M strategies are mainly related to how platforms manage application execution; and human aspects are worker characteristics. Each dimension is composed of a set of factors. Figure~\ref{fig1} provides an overview of the framework.

Considering their relations, it is clear that the dimensions are not independent. The definition of D\&M strategies is affected by both the QoS requirements and the human aspects. QoS requirements reflect requester objectives that should guide the design of suitable D\&M strategies. Human aspects, in turn, consist in workers' characteristics that delimit a restriction space where D\&M strategies may act aiming at optimizing QoS requirements. D\&M strategies generate a final output whose quality is a measure of their capacity of optimizing QoS requirements taking into account human aspects. In the following we detail our framework by discussing the theories that support its dimensions and their factors.

\subsection{QoS requirements}

The QoS requirements dimension encompasses a set of quantitative characteristics that indicate requesters' objectives and how they evaluate application effectiveness. QoS requirements have been mainly addressed in two distinct areas: process management for organizations~\cite{Aalst02workflowmanagement}, and QoS for software systems~\cite{Denning2003}. Based on the literature from these areas, we define QoS requirements in terms of the following factors:

\begin{itemize}
\item \textit{Time} refers to the urgency to transform an input into an output. It includes response time, and delays on work queues, and a time limit (deadline) for generating an output;
\item \textit{Cost} refers to expenditure on the application execution. It is usually divided into enactment and realization cost. Enactment cost concerns expenditure on D\&M strategies, and realization cost, expenditure on application execution.
\item \textit{Fidelity} reflects how well a task is executed, according to its instructions. Fidelity cannot be described in a universal formula and it is related to specific properties and characteristics that define the meaning of ``well executed''~\cite{Cardoso:2002}. It is a quantitative indicative of accuracy and quality.
\item\textit{Reproducibility} refers to obtain similar output when the application is executed at different times and/or by different group of workers taken from the same population~\cite{Paritosh:2012}.
\item \textit{Security} relates to the confidentiality of application tasks and the trustworthiness of resources that execute them.
\end{itemize}

\subsection{D\&M strategies}

Based on application design and management methodologies in machine-based computation~\cite{Cardoso:2002,Yu:2005,cirne:2003} and in organizations~\cite{Kumar:2002,Kiepuszewski:2000}, we define five factors for the D\&M strategies dimension: application composition, incentives and rewards, dependency management, task assignment, output aggregation, and fault prevention and tolerance.

\textit{Application Composition.} It consists of two major activities: problem decomposition and application structuring. Problem decomposition includes the following decisions: $1)$ tasks granularity, e.g., generating fewer task that require more effort to be executed (coarse-grained) or generating many small tasks that require less effort to be executed (fine-grained); $2)$ worker interfaces for the tasks, i.e., the interface design of the web page that shows the instructions of the work to be done.

\begin{figure}[!h]
\centering
\includegraphics[width=0.35\textwidth]{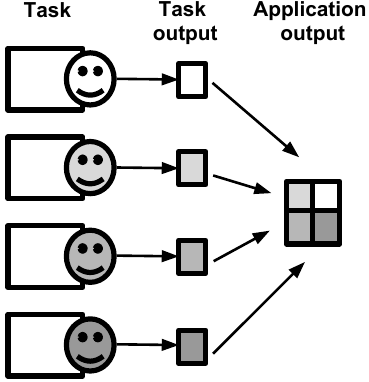}
\caption{Bag of human computation tasks.
Example of a bag-of-tasks application structuring.}
\label{fig2}
\end{figure}

\begin{figure}[!h]
\centering
\includegraphics[width=0.35\textwidth]{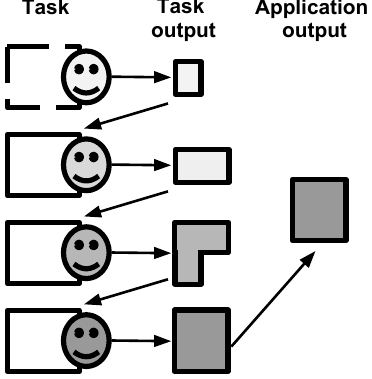}
\caption{Workflow of human computation tasks.
Example of a workflow application structuring.}
\label{fig3}
\end{figure}

Application structuring consists in how to compose the application considering possible dependencies between its tasks. As exemplified in Figures~\ref{fig2} and~\ref{fig3}, the two major application structuring patterns are \textit{bag-of-tasks} and \textit{workflow}. Bag-of-tasks applications are composed of a set of independent tasks. For example, a group of Human Intelligence Tasks (HITs) in MTurk platform~\cite{Little2010}. Workflow applications, in turn, are composed of a set of tasks organized in a sequence of connected steps~\cite{Christoph:2012}. Each step is composed by one or more tasks. Independent tasks are usually grouped in the same workflow step. Interdependent tasks, in turn, constitute different workflow steps.

\textit{Incentives and Rewards.} Incentive are put in place when the participants exhibit distinct objectives and the  information about them are decentralized~\cite{laffont2009theory}. In human computation systems, requesters and workers may have different interests. For example, some  workers may be interested in increasing their earnings, while requesters are interested in which tasks are performed with greater accuracy. Incentive strategies are used to align the interests of requesters and workers~\cite{Scekic:CommunACM:2013}. They are usually put in place to incentivize workers to exhibit a behavior and achieve a performance level desired by the requester, which includes executing more tasks, improving accuracy and staying longer in the system. Incentives can be broadly divided into non-monetary and monetary. Examples of non-monetary incentives are badges provided as a recognition for workers' achievements, and rank leaderboard for the workers to gauge themselves against peers. Monetary incentives are usually associated with a reward scheme, which defines the conditions for a worker to be rewarded, e.g., providing an output identical to the majority of other workers who perform the task. Game theory is an important theoretical guide to incentivize workers' engagament and effort in human computation systems~\cite{Arpita:2013,Jain:2009}.

\textit{Dependency Management.} It focuses on the coordination between interdependent tasks. A framework of dependencies between tasks in human computation is presented by Minder et al.~\cite{Minder2011}. It is mainly based on the Coordination Theory~\cite{Malone1994}. Dependencies among tasks can be broadly divided into four levels: serialization, visibility, cooperation, and temporal. Serialization dependencies specify whether tasks in the application require a serial execution. Such dependencies are usually defined in application structure by routing operations, such as: sequence, parallelism, choice and loops. Visibility dependencies define whether the work performed in a task must be visible to the other tasks (e.g., when a task updates a global variable). Cooperation dependencies, in turn, define which tasks hold a shared object at each time and can perform operations on it without restriction. Finally, temporal dependencies specify whether a set of tasks must perform operations in a particular temporal dependency.

\textit{Task Assignment.} It defines how to choose which worker will execute a task. The strategies can be broadly divided into scheduling, job board, and recommendation. Scheduling is a push assignment; workers receive and execute tasks assigned to them. Scheduling strategies assign tasks to workers trying to optimize one or more QoS requirements. It is usually based on application and/or workers information. Job board, in turn, is a pull assignment; workers use search and browser functionalities to choose the tasks they want to execute. It allows workers to select those tasks they expect to enable them to maximize their metrics, such as: earnings, preferences, and enjoyment. Finally, Recommendation is a hybrid assignment; workers receive a set of tasks and they choose which of them they want to perform. Recommendation is mapped into scheduling when the amount of tasks recommended is $1$, and it is mapped into job board when all tasks are recommended.

\textit{Output Aggregation.} It is concerned with aggregating sets of individual task outputs to identify the correct output or to obtain a better output. It is interchangeably called judgment aggregation and crowd consensus. An aggregation function may be implemented in several ways and it may be executed by a machine or a human. A simple example is that of different task outputs that constitute different parts of the application output; thus, the aggregation is only a merge of task outputs. A more sophisticated aggregation function may ask workers to improve available task outputs and generate an application output. Note that output aggregation is an \textit{offline} procedure, i.e., it is executed after the outputs of all application tasks have already been obtained. There are also \textit{online} procedures which involve failure detection in each task output, as well as strategies to detect and manage cheating workers. We discuss online procedures in fault tolerance strategies.

\textit{Fault Prevention and Tolerance.} Faults are events which potentially may cause a failure. A failure is a malfunction or incorrect output. Thus, fault prevention consists in avoiding events which may cause failures and fault tolerance consists in identifying and managing failures after they occur. To analyze human error in human computation systems, we join together human error concepts from Human Error Theory~\cite{James:HumanError:1990} and concepts related to the implementation of fault prevention and tolerance in computing system from Human-in-the-Loop~\cite{Cranor:UPSEC:2008} and machine-based distributed systems~\cite{Jalote:FTD:1994} literatures.

To execute a task, a human first constructs a mental plan of a sequence of actions that ends with the conclusion of the task~\cite{James:HumanError:1990}. Three types of failures may occur in this process: \textit{mistakes}, \textit{lapses} and \textit{slips}. Mistake is a failure in constructing a suitable plan to execute the task, then the plan is not correct. Lapses and slips are failures in the execution of a correct plan. Lapses occur when the worker forgets to perform one action of the plan. Finally, slips occur when the worker performs incorrectly an action of the plan. A diversity of faults can generate such failures, for example: lack of knowledge or time to execute the task, and stochastic cognitive variability such as variability of attention.

Fault prevention strategies usually focus on methodologies for design, testing, and validation of tasks instructions, and testing resources capabilities. Fault tolerance, in turn, consists of four phases: failure detection, damage confinement, failure recovery, and fault treatment. Failure detection consists of identifying the presence of a failure, e.g., identifying that a task output is incorrect. Damage confinement aims at determining the boundaries of failures and preventing their propagation, e.g., identifying which tasks outputs are incorrect and preventing that other tasks make use of these outputs. Failure recovery tries to bring the flow of execution to a consistent state, e.g., re-executing tasks that produced incorrect outputs. Finally, fault treatment involves treating faults to prevent the occurrence of new failures.

\subsection{Human aspects}

In our context, human aspects are human beings characteristics that determine the way they perform tasks. These aspects have been widely addressed in Psychology studies. They can be broadly divided into the following factors: cognitive system~\cite{Simon:1990,Sweller:1998}, motivation~\cite{Maslow:1943}, preferences~\cite{Kapteyn:1978}, social behavior~\cite{Alexander:1974}, emotion~\cite{Gross:1998,Dolan:2002}, individual differences~\cite{Parasuraman:2012,Stanovich:1998}, and intraindividual changes~\cite{Ram:2005}.

\textit{Cognitive system.} Its function includes several processes of task execution, such as information processing, understanding, and learning. It specifies processes' organization on long-term and short-term memory. Long-term memory is where knowledge is stored. In turn, short-term memory is a working memory used to process information in the sense of organizing, contrasting, and comparing~\cite{Simon:1990}. Humans are able to deal with few items of information simultaneously in their working memory, and any interactions between items held in their working memory also require working memory capacity, reducing the number of items that can be dealt with simultaneously~\cite{Sweller:1998}. Cognitive overload occurs when tasks processing exceeds working memory capacity.

\textit{Motivation.} From the motivation theory viewpoint~\cite{Maslow:1943}, humans are guided by motivation impulses or goals, i.e., the desire to do/obtain new things and to achieve new conditions. Incentive studies explore the way such goals influence human behavior. Considering the self-determination theory~\cite{Deci:book:1985}, motivation is broadly divided into intrinsic and extrinsic. In task execution, intrinsic motivation may consist of workers internal desires to perform a particular task because it gives them pleasure or allows them to develop a particular skill. Extrinsic motivation, in turn, consists of factors external to the workers and unrelated to the task they are executing.

\textit{Preferences.} Humans exhibit personal preferences~\cite{Kapteyn:1978}. Such preferences are explained on the basis of two types of influences: their own past experiences and the experiences of others which are directly observable by them. As an example of workers preferences in task, consider the case where, after feeling bored several times when executing high-time-consuming tasks, workers always choose to execute only low-time-consuming tasks.

\textit{Social behavior.} Sociality means group/community organization to perform activities~\cite{coleman:book:2000}. In general, communities form and persist because the individual takes advantage of them and thereby serve their interests. Sense of Community theory suggests that members develop sense of community based on membership, influence, integration and fulfillment of needs, and shared emotional connection~\cite{mcmillan:JCP:1996}. This behavior may influence the way community members behave and execute tasks in the system.

\textit{Emotion.} Emotion can be defined as a human complex psychological and physiological state that allows humans to sense whether a given event in their environment is more desirable or less desirable~\cite{Dolan:2002}. Emotion concerns, for instance, mood, affection, feeling and opinion. It interacts with and influences other human aspects relevant to task execution effectiveness. For example, it influences cognitive system functions related to perception, learning and reasoning~\cite{Dolan:2002}.{\pagebreak} 

\textit{Intraindividual variability and change.} Humans show intraindividual variability and change~\cite{Ram:2005}. Intraindividual variability is a short-term or transient fluctuation characterized by factors as: wobble, inconsistency, and noise. Intraindividual change is a long-term lasting change as a result of learning, development, or aging.

\textit{Individual differences.} Humans show variability between themselves in several factors~\cite{Parasuraman:2012,Stanovich:1998}, such as decision making and performance. In this study we focus mainly on individual differences in terms of the following three human competencies: knowledge, skills, and abilities. Knowledge refers to an organized body of information applied directly in the execution of a task. Skill refers to the proficiency in tasks execution, usually measured qualitatively and quantitatively. Abilities are those appropriate on-the-job behaviors needed to bring both knowledge and skills in task execution.

\section{Instantiating the framework}

Now we turn to map the literature on human computation by using our theoretical framework. This is done through a literature review focused on analyzing: a) how human computation studies on D\&M strategies have dealt with human aspects discussed in our framework to satisfy the QoS requirements of requesters; and b) what are the relevant results regarding these human aspects upon which future D\&M strategies decisions can be based. Throughout this section, for each D\&M factor, we discuss the human computation studies and extract the major implications for system design related to human factors.

\subsection{Application composition}
\label{subsubsec:composition}

Application composition consists of two major activities: task design and application structuring.

\textit{Task design} impacts on the ability of worker to complete tasks quickly, effectively, and accurately~\cite{Khanna:2010}. Poorly designed tasks, which show high cognitive load, can cause fatigue of workers, compromising their understanding of instructions, decreasing their productivity and increasing their errors~\cite{Kulkarni2012}. This usually occurs because of the limitations of the human working memory. This is the case of tasks where humans are asked to choose the best item among several options~\cite{Sun:Hcomp:2011,Venetis:2012}. To perform this task, humans compare the items and choose the best of them on their working memory. Such kind of task generates a cognitive load, which increases in proportion as the number of items to be compared increases. The higher the cognitive load, the higher error chances. {\color{black} Tasks can also be designed to motivate workers to put more effort in generating correct outputs. Huang et al. show that one way to achieve it is to tell workers  explicitly that their work will be used to evaluate the output provided by other workers~\cite{Huang:2013}.}

\textit{Application structuring} studies can be broadly divided into static workflows and dynamic workflows. Example of static workflow composition is that used in Soylent~\cite{Bernstein:2010}. Soylent is a word processor add-in that uses MTurk's workers (mturk.com) to perform text shortening. Soylent implements the Find-Fix-Verify workflow, i.e., find areas of the text that can be shortened, fix and shorten the highlighted area without changing the meaning of the text, and verify if the new text retain its meaning. This task distinction captures human individual differences mainly in terms of the type of tasks workers want to perform, i.e., find, fix, or verify tasks. Static workflows can be optimized. Cascade~\cite{Chilton:2013} and Deluge~\cite{Bragg:2013} are examples of optimized workflows to taxonomy creation tasks.

Example of dynamic workflow composition is that used in Turkomatic~\cite{Kulkarni2012}. In Turkomatic, workers compose the workflow dynamically on a collaborative planning process. When a worker receives a task, she/he performs it only if it is simple to be executed; otherwise, the worker subdivides it into two other tasks. A problem occurs when workers generate unclear tasks that will be executed by other workers. Other approach is proposed by Lin et al.~\cite{Christopher:2012}. They assume that workflow composition results in different output quality when executed by different groups and they propose the availability of multiple workflows composition with independent difficulty level and dynamically switch between them to obtain higher accuracy in a given group of workers. Finally, Bozzon et al. propose a system that dynamically controls the composition and execution of the application, and reacts to some specific situations, such as: achievement of a suitable output and identification of a spammer~\cite{Bozzon:2013}. It allows the system to adapt the application to changes in workers characteristics and behavior.

\textbf{Implications for systems design.} We extract two major guidelines from this discussion: $1)$ application designers must avoid task cognitive overload, in what requiring a small amount of specialized ability, skill and knowledge in a same task can contribute; $2)$ given that workers in a platform display individual differences, application designers can take advantage of worker diversity by developing applications with different types of tasks, each type requiring a different skill; this may be done either by defining a static composition of tasks that require different skills or by using dynamic composition to adapt to different groups of workers.

\subsection{Incentives and rewards}

Incentive and reward schemes have been designed to incentivize specific behaviors and maximize requesters' QoS requirements~\cite{Singla:www:2013,Singer:www:2013,Scekic:CommunACM:2013}. Unfortunately, there is no consensus in the literature on the effect of incentives on worker behavior. Its effect seems to vary with the type of task. In some tasks, increasing financial incentives allows one to increase the amount of workers interested in executing the task~\cite{Barowy:2012,Mason:2009,Archak:2010}, but not necessarily the quality of the outputs~\cite{Mason:2009}. In other tasks, quality may be improved by increasing intrinsic motivations~\cite{Rogstadius:ICWSM:2011}. Incentives also relates to other aspects of task design, for example, some studies show that incentives work better when they are placed in the focal position in task worker interface~\cite{Chandler:HCOMP:2011}, and task fidelity is improved when financial incentives are combined with a task design that asks workers to think about the output provided by other workers for the same task~\cite{Shaw:CSCW:2011}.

Besides defining right incentives, requesters must also define a suitable reward scheme. Witkowski et al. analyze a scheme that pays workers only if his/her output agrees with those provided by others, and that penalizes with negative payment in the case his/her output disagrees. They show that such scheme acts as a self selection mechanism that makes workers with lower quality choose not to participate~\cite{Witkowski:2013}. Rao et al. show that a reward scheme that informs workers that they will be paid if their outputs are similar to the majority motivates workers to reflect more on what other workers would choose. This generates a higher percentage of correct outputs and the obtained outputs are closer to the correct one~\cite{Rao:2013}. Huang et al. analyze three schemes in a group of workers~\cite{Huang:CHI:2013}: (i) individual, in which reward depends only on worker performance; (ii) teamwork, in which workers in the group are paid similarly based on the average of their performance; and (iii) competition, in which workers in the group are opponents and they are paid based on their differences of performance. The effectiveness of these schemes tends to vary with other application settings such as social transparency.

\textbf{Implications for systems design.} We extract two major guidelines from this discussion: $1)$ given that the effect of incentives on intrinsic and extrinsic motivations appears to be different in different types of tasks, designers must test how combinations of such incentives contribute towards the desired quality in their specific context; $2)$ task peformance can be improved by using incentives that motivate workers to reflect more on what output other workers would provide for the same task.

\subsection{Task assignment}

We broadly divided task assignment strategies into scheduling, job board, and recommendation.

\textit{Task Scheduling} strategies try to optimize some QoS requirements by exploiting information about the affinity between tasks and workers. Scheduling strategies in human computation literature have considered several human aspects. Some strategies consider workers' emotional states allocating tasks that are appropriate to current worker's mood~\cite{Morris:2011}. {\color{black}Heidari and Kearns take into account task difficulty and workers abilities~\cite{Heidari:2013}. They  analyze the case in which workers can decide between execute one task or forward it. When one worker decides to forward a task, it will be scheduled to a more qualified worker. It generates a forwarding structure on task scheduling that improves outputs' quality. Waterhouse proposes a strategy based on a probabilistic metric of mutual information between workers. The strategy tunes the assignment of tasks to the workers that will increase the amount of information gain~\cite{Waterhouse:2013}.}

Other thread of scheduling strategies is inspired by the hierarchical structure in today's organizations, exploring individual differences~\cite{Noronha:2011,Schallwwwj:2012}. They consider working teams and roles, such as supervisors and workers. Tasks are first assigned to supervisors, who assign them to workers in their team taking into account the qualification and skills of each worker. Skill-based scheduling considers different workers qualification levels and that qualification level increases in the proportion that workers adequately perform more tasks~\cite{Satzger:2011}. There are also approaches that use information and contents liked by the worker on social networks to automatically match workers preferences and task requirements~\cite{Difallah:www:2013}.

\textit{Job Board} strategies are used mainly in online labor market platforms, where tasks are usually made available together with their rewards to workers in boards~\cite{Chilton:HCOMP:2010}. {\color{black}Job boards allow workers to choose tasks that fit their preferences~\cite{Lee:2013}; thus, task instructions must be clear and attractive to workers~\cite{Jason:2013}}. Requesters define job parameters so as to address workers interests and make their tasks attractive to workers. For example, AutoMan adjusts tasks' price and tasks' allocation time to motivate more workers to try executing them~\cite{Barowy:2012}. By adjusting these parameters, AutoMan tries to optimize QoS requirements addressing workers' time and financial incentives. Toomim et al. propose a method to characterize workers preferences for some interfaces and tasks~\cite{Toomim:CHI:2011}. This information is used in future task design.

\textit{Task Recommendation} strategies recommend tasks to workers according to some affinity criteria~\cite{Ambati:HCOMP:2012}. The platforms oDesk (odesk.com) and Elance (elance.com) are based on job boards, but they also make use of a recommendation system to inform workers about new jobs that match their skills. We are not aware of studies that evaluate the effectiveness of such functionalities on these platforms. {\color{black}Yi et al. propose a matrix completion approach to infer workers preferences in pairwise comparison tasks~\cite{Yi:2013}. The method may be useful to improve task recommendation strategies.}

In job board and tasks recommendation approaches, it is also required a mechanism that allows requesters to choose which of the candidate workers will perform the task or which solution will be paid. In research and practice, three strategies that explore these dimensions are: \textit{auction}, in which the task is allocated to the worker that charges the lowest value (e.g., odesk.com); \textit{challenge}, in which all workers perform the available tasks, but only the best solution is paid (e.g., topcoder.com~\cite{Archak:2010}); and \textit{registration order}, in which the task is allocated to the first worker that signed up to run it (e.g., mturk.com~\cite{Chen:chiHcomp:2011}). To the best of our knowledge, no study was conducted to compare the performance of these approaches and to indicate in what situations each should be used.

\textbf{Implications for systems design.} Two basic guidelines to highlight in this context are: $1)$ given that most of platforms are based on job boards and that in this environment the effectiveness of a task relies on its ability to gain attention of suitable workers, requesters must provide task descriptions with information that allows workers to easily identify if the task match their skills, preferences and interests; $2)$ requesters must avoid generating too restrictive tasks in order to take advantage of the diversity and larger quantities of workers that job boards and recommendation strategies give access to.

\subsection{Dependency management}

We focus on analyzing the dependency management strategies that take into account human aspects, while ensuring temporal, serialization, visibility, and cooperation dependencies. Most studies address only temporal dependencies, in which a set of tasks must be performed in a particular order (e.g.,~\cite{Noronha:2011,Little2010}) or in a synchronous way (e.g.,~\cite{Kearns:2012,Mao:2011}).

\textit{Serialization} dependency studies in human computation have focused maily on applications with loop or without loop. Example of human computation applications without loops are those that deal with planning activities~\cite{LawZ11}. Such applications usually include a sequence of steps such as: decomposition of the problem into small tasks, execution of each task and aggregation of these partial task outputs to obtain the final output as a result to the problem. This is the case of Turkomatic~\cite{Kulkarni2012}, CrowdForge~\cite{Kittur2011}, CrowdPlan~\cite{LawZ11}, and combination of creative design~\cite{Yu:CHI:2011}. Application with loops, in turn, include some iterative processes~\cite{Little2010} as in Find-Fix-and-Verify~\cite{Bernstein:2010} and Iterative Improvement~\cite{dai:AAAI2010}.

\textit{Visibility} dependency is common in working group. It usually needs a shared environment that unobtrusively offers up-to-date group context and explicit notification of each user's action when appropriate. Mao et al.~\cite{Mao:2011} and Zhang et al.~\cite{Zhang:2012} address visibility dependencies in human computation applications. In their studies, workers try to achieve a global goal. This goal can be, for example, a collaborative itinerary planning~\cite{Zhang:2012}, a collaborative graph coloring~\cite{Mao:2011}. In these cases, workers can see outputs generated to correlated tasks. Such type of task is usually related to agreement or consensus and the visibility decision may impact both worker behavior as well as the time required to obtain an output~\cite{Kearns:2012}.

\textit{Cooperation} dependencies are also related to working group. Mobi~\cite{Zhang:2012} and TurkServer~\cite{Mao:2011} allow one to implement applications that contain cooperative tasks. Zhang et al. show that the unified view of the task status allows workers to coordinate and communicate more effectively with one another, allowing them to view and build upon each other's ideas~\cite{Zhang:2012}. Platforms such as MTurk maintain workers invisible and not able to communicate with each other. Turkopticon is a tool that allows one to interrupt such invisibility, making possible workers to communicate among themselves~\cite{Irani:2013}.

\textbf{Implications for systems design.} The major guideline regarding dependency management is that designers must consider that some degree of visibility and communication between workers may be positive for application performance in terms of time and accuracy. It seems that workers should be allowed: $1)$ to see the status of tasks that are interdependent with his/her task in order to synchronize its execution with any global task constraint; and $2)$ to communicate with other workers that are executing tasks interdependent with his/her task in order to improve cooperation.

\subsection{Output aggregation}

There are a range of output aggregation strategies in human computation, most of them are already discussed by Law and von Ahn~\cite{Law:Book:2011} and Nguyen et al.~\cite{Nguyen:WISE:2013}. We focus on discussing studies that account for human aspects.

An example of comparable outputs aggregation strategy is majority vote~\cite{Barowy:2012,Little2010}, in which the output of the application is the most frequent task output. This strategy assumes that the majority of the workers assigned to any task are reliable. Majority vote does not perform properly when the error chance in task execution is high. Sheng et al. investigate the impact of the number of task executions on output accuracy and shows that quality of the output is improved by using additional workers only when the workers accuracy is higher than $0.5$~\cite{Sheng:KDD:2008}. They propose a repeated-labeling technique that selects data points for which application quality should be improved by the acquisition of multiple task outputs.

Diverse studies have been devoted to aggregating a set of outputs and obtaining an accurate output taking into account workers expertise and task characteristics. Whitehill et al. consider that an output accuracy depend on the difficulty of the task and expertise of the worker~\cite{whitehill:ANIPS:2009}. They proposed Generative Model of Labels, Abilities, and Difficulties (GLAD) which estimates these parameters using Expectation Maximization (EM) probabilistic models and evaluate tasks output in a way that experts' outputs count more. Hovy et al. propose Multi-Annotator Competence Estimation (MACE) which uses EM to identify which annotators are trustworthy and considers this information to predict outputs~\cite{Hovy:NAACL-HLT:2013}. Wang et al. propose a recursive algorithm to improve the efficiency of compute EM models in these contexts~\cite{Wang:2013}. {\color{black}Dalvi et al. propose a technique to estimate worker reliabilities and improve output aggregation~\cite{Dalvi:2013}. The strategy is based on measuring agreement between pairs of workers.}

Another output aggregation challenge arises in unstructured outputs, e.g., open-ended~\cite{Bernstein:2010,Sun:Hcomp:2011} {\color{black}and image annotation tasks~\cite{Salek:2013}}. In this case, a way to find the best output is to apply a vote-on-the-best strategy in which workers evaluate the quality of each output or they choose which of them exhibits the highest quality~\cite{Kulkarni2011b}. It exploits individual differences, given that some workers are better at identifying the correct outputs than producing them themselves~\cite{Sun:Hcomp:2011}. When the set of options is too large, it may be difficult for workers choose the best item. An alternative in this case is to develop a second human computation application in which few items are compared in each task and the best item is chosen by tournament (e.g., ~\cite{Sun:Hcomp:2011,Venetis:2012}). Other peculiarity of unstructured outputs is that even poor outputs may be useful. For example, other workers can aggregate such poor outputs, and generate a new better output~\cite{Kulkarni2012}. {\color{black}The quality of the aggregation can also be improved by using estimations of the difficulty level of tasks and skills of workers~\cite{Salek:2013}.}

\textbf{Implications for systems design.} When developing output aggregation strategies, designers must weigh at least three parameters that impact on the quality of the final output: $1)$ task cognitive load; $2)$ the amount of different workers that provided task outputs, i.e., redundancy degree; and $3)$ the accuracy of each worker that provided the outputs. As in the literature, the value of each of parameters can be obtained in a statistical evaluation, considering that the accuracy of the final output tends to be higher with more accurate estimation of these parameters.

\subsection{Fault prevention and tolerance}

The prevention of faults in task instructions can be done by using offline and/or online pilot tests~\cite{Chen:chiHcomp:2011}. Offline tests are conducted with accessible people that can provide feedback on issues and improvements in the tasks instructions. Online tests, in turn, are driven onto a platform, and they are more realistic than offline tests. In this case, workers may not be accessible to provide feedback about the task instructions, but their outputs can be analyzed to identify problems.

The prevention of undesired workers is usually done by using qualification tests~\cite{Chen:chiHcomp:2011}. They consist in requiring the execution of a gold standard test that certifies whether the worker has the skills and qualifications required to perform application tasks. Only workers who perform accurately are considered qualified. A downside of this approach is not considering changes in workers behavior after executing the test. Malicious workers usually change their behavior over time~\cite{Vuurens-CIR2011}. CrowdScape is a system that allows requesters to select workers based on both task output quality and workers' behavioral changes~\cite{Rzeszotarski:UIST:2012}.

Studies also have been devoted to fault tolerance which consists in four phases: failure detection, damage confinement, failure recovery, and fault treatment.

\textit{Failure detection} has been made by using: $1)$ conformance voting, which allows one to detect poorly executed tasks; and $2)$ timing, which allows one to detect worker evasion, i.e., the worker is assigned to perform a task, but gives up executing and do not deallocate the task. In conformance vote, one worker or a group of workers evaluate whether a task output is correct. When the output is not correct, the task needs to be re-executed by another worker~\cite{Kulkarni2012}. Timing, in turn, defines a maximum time that a task can remain allocated to a worker; it is expected that a worker provides an output up to this time. If that time expires without an output being provided, the task is deallocated and made available to another worker~\cite{Bernstein:2010}.

\textit{Damage Confinement} is usually made by using error recovery techniques in each task or workflow step. It prevents that damages propagate to the next workflow step. This propagation occurs, for example, in workflow derailment problems~\cite{Kulkarni2011b}, which arises when an error in the task executions prevents the workflow conclusion.

\textit{Failure Recovery} has been made by using majority voting, alternative, and human assessment. These strategies exploit human individual differences by using redundancy of workers. If different and independent workers provide the same output to a task, it increases the confidence that the task is being performed in accordance with its instructions~\cite{Kochhar:2010}. In majority voting, several workers perform the same task in parallel and the most frequent output is considered correct. In alternative strategies, in turn, a worker executes the task and, if an error occurs, the task is executed again by another worker~\cite{Kulkarni2012}. {\color{black}In these redundancy-based strategies, the impact of the redundancy degree on output accuracy is highly dependent on the type of task. Increasing the redundancy of workers does not necessarily increase the confidence that the correct output will be obtained~\cite{Amir:2013}. Furthermore, the perception of redundancy by the workers may have a negative effect on their motivation and work quality. The more co-workers working in the same task are perceived by workers, the lower their work quality~\cite{Kinnaird:2013}. This occurs because workers demotivate thinking that their effort does not count for much.

Finally, in human assessment strategies, the outputs generated by a worker are evaluated by others. This can be implemented in two ways: arbitration and peer review. In arbitration, two workers independently execute the tasks and another worker evaluates their outputs and solve disagreements. In peer review, the output provided by each worker is reviewed by another worker. Hansen et al. show that in text transcribe tasks, the peer review strategy is significantly more efficient, but not as accurate for certain tasks as the arbitration strategy~\cite{Hansen:2013}.}

\textit{Fault Treatment} has been made by fixing problems in task design, and by eliminating or reducing the reputation of unskilled or malicious workers. For example, TopCoder~\cite{Archak:2010} maintains a historical track of the number of tasks each worker chooses to execute, but did not conclude. This track is used to estimate the probability that the worker chooses tasks and do not execute it. Ipeirotis et al. propose to separate systematic errors from bias due to, for example, an intraindividual variability such as distraction~\cite{Ipeirotis:HCOMP2010}. This distinction allows also a better estimation of accuracy and reputation of the worker. Such estimation may be used to prevent assigning to a worker tasks for which he/she is not qualified or that he/she will not complete the execution. Another important aspect in fault treatment is to provide feedback to workers about his/her work~\cite{Dow:CSCW:2012}. It helps workers to learn how to accurately execute the task (intraindividual changes) and avoid errors due to lapses (intraindividual variability)~\cite{Ipeirotis:HCOMP2010} .

\textbf{Implications for systems design.} The three major guidelines extracted from this discussion are: $1)$ designers must test the task worker interface and check workers skills/reputation; to this end pilot tests and qualification tests can be applied; $2)$ redundancy is the basis of fault tolerance strategies, but requesters must generate tasks that maximize the number of workers capable of executing it, increasing the potential of redundancy of the task; and $3)$ requesters must provide workers assessment and feedback in order to allow them to learn from tasks they perform incorrectly.\vspace*{-5pt} 

\section{Challenges and perspectives}

In the last section, we analyzed the human computation literature and its implication for design in the light of our theoretical framework. Now we turn to discuss challenges and perspectives in D\&M strategies. Although our list is by no means exhaustive, it offers examples of topics in need of further work and directions that appear promising.

\subsection{Relations between dimensions of the framework}

Table~\ref{tab1} synthesizes the contributions on the relationships between D\&M strategies and human aspects identified in the last section. As shown, there are several relationships for which we could not find any study. This state of affairs indicates a large amount of research still to be conducted after mapping the impact of human aspects on D\&M effectiveness. Two other issues that still require further understanding are: $1)$ adequate combinations of D\&M strategies; and $2)$ the impact of D\&M strategies on workers' cognition and behavior.

\begin{sidewaystable}
\caption{\bf Human aspects factors addressed in D\&M strategies}\label{tab1}
\begin{tabular}{@{\extracolsep\fill}ccccccc@{\extracolsep\fill}}
\hline
& {\textbf{Application}} & {\textbf{Incentives}} & {\textbf{Dependency}}& {\textbf{Task}}&{\textbf{Output}}&{\textbf{Fault}}\\
{} & {\textbf{composition}} &{\textbf{and rewards}} & {\textbf{management}}& {\textbf{assignment}}&{\textbf{aggregation}}&{\textbf{tolerance}}\\\hline\\[-10pt]

{Cognitive System} &\cite{Venetis:2012,Kulkarni2012,Khanna:2010,Sun:Hcomp:2011}&&&\cite{Heidari:2013}& \cite{whitehill:ANIPS:2009,Salek:2013}&\cite{Ipeirotis:HCOMP2010}\\[2pt]\hline\\[-10pt]

{Motivation} &&\cite{Archak:2010,Barowy:2012,Mason:2009,Huang:2013,Witkowski:2013, Chandler:HCOMP:2011,Singla:www:2013,Singer:www:2013,Rao:2013}& &\cite{Barowy:2012}&& \cite{Mason:2009,Rogstadius:ICWSM:2011,Kinnaird:2013} \\[2pt]\hline\\[-10pt]

{Preferences} &\cite{Bernstein:2010} && &\cite{Toomim:CHI:2011,Chilton:HCOMP:2010,Ambati:HCOMP:2012,Lee:2013, Jason:2013}&&\\[2pt]\hline\\[-10pt]

{Social behavior} && \cite{Shaw:CSCW:2011,Huang:CHI:2013}& \cite{Kearns:2012,Zhang:2012,Mao:2011,Irani:2013} &\cite{Difallah:www:2013} &&\cite{Barowy:2012}\\[2pt]\hline\\[-10pt]

{Emotion} & &&& \cite{Picard:2003,Morris:2011}&& \cite{Ipeirotis:HCOMP2010}\\[2pt]\hline\\[-10pt]

{Individual differences} &\cite{Bernstein:2010,Kulkarni2012,Christopher:2012} && &\cite{Noronha:2011,Schallwwwj:2012,Archak:2010,Waterhouse:2013} &\cite{Sheng:KDD:2008,whitehill:ANIPS:2009,Hovy:NAACL-HLT:2013,Kulkarni2012, Bernstein:2010,Sun:Hcomp:2011,Dalvi:2013}& \cite{Kulkarni2012,Bernstein:2010,Rzeszotarski:UIST:2012,Kulkarni2011b, Hansen:2013}\\[2pt]\hline\\[-10pt]

{Intraindividual Variability} & && & && \cite{Ipeirotis:HCOMP2010,Dow:CSCW:2012,Rzeszotarski:UIST:2012}\\[2pt]\hline\\[-10pt]

{Intraindividual changes} &&&&\cite{Satzger:2011} &&\cite{Dow:CSCW:2012,Vuurens-CIR2011}\\[2pt]\hline\\[-10pt]
\end{tabular}
\end{sidewaystable} \clearpage

It is intuitive that one D\&M strategy may impact on the effectiveness of another. For example, by generating a fine-grained application composition to account for the human cognitive system, one may generate undesired effects: $1)$ designing tasks too susceptible to cheater workers, which reduces the effectiveness of fault tolerance strategies; or $2)$ generating a large number of tasks with a too high number of dependencies among them, which may reduce parallelism in task execution and impact on dependency management. More empirical research on how to adequately combine D\&M strategies in distributed human computation is still required.

Besides the requesters' perspective that tries to understand how to take advantage of human aspects to achieve QoS requirements, studies must also identify possible side-effects of the strategies on workers cognition and behavior. Two cognitive effects that may be relevant to consider are: \textit{Framing effect} -- workers may generate different outputs based on how a task is designed--, and \textit{Hawthorne effect} -- workers may alter their behavior when they know that they are being observed. Two behavioral effects are collusion, an agreement between workers to act similarly (e.g., planning collusion against requesters which submit poorly designed tasks~\cite{Kulkarni2012}), and sabotage, workers change their behavior to take advantage of the system (e.g., inhibiting competitors in a ``maximum observability'' audition~\cite{Archak:2010}). Also, there is room for studies focused on workers and on the fair relationship between workers and requesters~\cite{Silberman:2010}. 

\subsection{Exploring the Interdisciplinarity of D\&M strategies improvement}

\textit{Application composition.} The main human aspects factors that have been addressed in application composition are cognitive system and motivation/incentives. By taking into account such factors in the context of task execution, human computation application composition is clearly related to the disciplines: \textit{ecological interface}~\cite{Rasmussen:1989}, and \textit{goal setting}~\cite{Locke:2002}. Ecological interface principles are grounded on how the human cognitive system works and its effects on information understanding and processing. Such principles may support the development of task designs to avoid cognitive overload and improve task execution effectiveness. Goal setting studies, in turn, may help better defining both task instructions and the way their outputs will be evaluated by the requester. Knowledge of such topics and reasoning about their relationships to human computation can help in the formulation of new strategies.

\textit{Task assignment.} Studies on task assignment have mainly taken into account: preferences and individual differences. Two other disciplines that take into account these aspects in task assignment are \textit{person-job fit}~\cite{edwards1990person} and \textit{achievement motivation}~\cite{nicholls1984achievement}. The domain of person-job fit research consists on tasks characteristics, worker characteristics, and required outcomes. It emphasizes the fit and matching of workers and tasks in the prediction of both worker and organizational outcomes. Achievement motivation is a motivation for demonstrating high rather than low ability. This motivation influences the tasks a human chooses to perform, i.e., his/her preferences. According to this concept, individuals select tasks they expect to enable them to maximize their chances of demonstrating high ability and avoiding demonstrating low ability. These concepts may inspire tasks scheduling and task recommendation strategies in human computation.

\textit{Dependency management.} Ensuring tasks dependencies and still extracting the greatest potential (optimizing QoS requirement) of a crowd of workers is one of the main challenges of dependency management strategies. Similar challenge has been addressed in at least two other disciplines: \textit{work teams}~\cite{Hackman:workteams:1987} and \textit{Groupware}~\cite{Karsenty:1993}. Both disciplines focus on group behavior and performance. Work team studies usually focus on group work in an organization not necessarily performed through a computer system. Groupware is generally associated with small groups working collaboratively through a computer system. Experiences on how human aspects are addressed in these disciplines may inspire solutions that consider these factors in human computation.{\pagebreak} 

\textit{Output aggregation.} Two important areas related to output aggregation are \textit{Judgment Aggregation}~\cite{dietrich:2007} and \textit{Social choice theory}~\cite{Taylor:Book:2008}. Judgment aggregation is the subject of a growing body of work in Economics, Political science, Philosophy and related disciplines. It aims at aggregating consistent individual judgments on logically interconnected propositions into a collective judgment on those propositions. In these situations, majority voting cannot ensure an equally consistent collective conclusion. Social choice theory, in turn, is a theoretical framework for analysis of combining individual preferences, and interests to reach a collective decision or social welfare. According to this theory any choice for the entire group should reflect the desires/options of the individual to the extent possible. The studies that have been conducted in these disciplines seem to be related to human computation output aggregation~\cite{Andrew:2013,Pettit:PhilosophicalIssues:2011}. A better mapping of their similarity and differences may help in the reuse and development of new output/judgment aggregation strategies.

\textit{Fault prevention and tolerance.} Besides preventing and tolerating faults, one should also consider how to evaluate system QoS in the presence of human faults. For example, fault tolerance is mainly based on task redundancy, but defining the appropriate level of redundancy is a challenging task. Maintaining a low level of redundancy may not recover failures and maintaining a high level of redundancy can lead to a high financial cost or high volunteer effort to run the entire application. This kind of study has been conducted in other disciplines such as: \textit{human aspects evaluation}~\cite{Whitefield:BehaviourInformationTechnology:1991} and \textit{performability}~\cite{misra2008handbook}. Human aspects evaluation is an assessment of the conformity between the performance of a worker and its desired performance. Performability, in turn, focuses on modeling and measuring system QoS degradation in the presence of faults. Experiences on performability and human aspects evaluation may be useful to address QoS requirements in the presence of worker faults.\vspace*{-5pt} 

\section{Conclusions}

In this paper, we analyzed the design and management of distributed human computation applications. Our contribution is three-fold: $1)$ we integrated a set of theories in a theoretical framework for analyzing distributed human computation applications; $2)$ by using this theoretical framework, we analyzed human computation literature putting into perspective the results in this literature on how to leverage human aspects of workers in D\&M strategies in order to satisfy the QoS requirements of requesters; and $3)$ we highlighted open challenges in human computation and discussed their relationship with other disciplines from a distributed systems viewpoint.

Our framework builds on studies in different disciplines to discuss advances and perspectives in a variety of immediate practical needs in distributed human computation systems. Our literature analysis shows that D\&M strategies have accounted for some human aspects to achieve QoS requirements. However, it also shows that there are still many unexplored aspects and open challenges. Inevitably, a better understanding of how humans behave in human computation systems and a proper delimitation of the human aspects involved is essential to overcome these challenges. We hope our study inspires both discussion and further research in this direction.\vspace*{-5pt} 

\section*{Competing interests}

The authors declare that they have no competing interests.\vspace*{-5pt} 

\section*{Authors' contributions}

LP, FB, NA, and LS jointly designed the theoretical framework used to contextualize and discuss the literature in this survey. LP drafted most of the manuscript and conducted the bulk of the review of the literature. FB, NA and LS did a smaller portion of the review of the literature and revised the manuscript in several interactions. All authors read and approved the final manuscript.\vspace*{-5pt} 

\section*{Acknowledgements}

Lesandro Ponciano thanks the support provided by CAPES/Brazil in all aspects of this research. Francisco Brasileiro acknowledges the support received from CNPq/Brazil in all aspects of this research.

\end{document}